\documentstyle[12pt]{article}
\newcommand{\be}{\begin{equation}}
\newcommand{\ee}{\end{equation}}
\begin{document}
\begin{titlepage}
\title{Generalized slow-roll inflation}
\author{Valerio Faraoni$^{1,2}$ \\ \\
{\small \it $^1$ Research Group in General Relativity (RggR)} \\ 
{\small \it Universit\'e Libre de Bruxelles,  Campus Plaine CP 231} \\
{\small \it  Blvd. du Triomphe, 1050 Bruxelles, Belgium}\\ \\
{\small \it $^2$ INFN-Laboratori Nazionali di Frascati, 
P.O. Box 13, I-00044 Frascati, Roma (Italy)} 
}
\date{} 
\maketitle   \thispagestyle{empty}  \vspace*{1truecm}
\begin{abstract} 
The slow-roll approximation to inflation is ultimately justified by the
presence of
inflationary attractors for the orbits of the solutions of the dynamical
equations in phase space. There are many indications that
the inflaton field couples nonminimally to the spacetime
curvature: the existence of attractor points for inflation
with nonminimal coupling is demonstrated, subject to a condition on the
inflaton potential and the value of the coupling constant.
\end{abstract}
\vspace*{1truecm} 
\begin{center} 
\end{center} 
\begin{center} 
\end{center}     
\end{titlepage}   \clearpage

A period of inflationary expansion of the early universe has come to be
widely accepted  in order to solve the horizon, flatness and monopole
problems that plague the standard big bang cosmology. As a bonus, 
inflation provides quantum fluctuations of the inflaton 
field as a much needed mechanism for  generating density perturbations, 
the seeds of structures observed in the universe today
\cite{KT,Lidseyetal}.

Most of the proposed inflationary scenarios are based on the slow-roll
approximation, which assumes that the  kinetic energy of the
inflaton is much smaller than its potential energy, and that the
cosmic expansion is nearly exponential; the slow-roll approximation allows
one to predict amplitudes and
spectra of scalar and tensor perturbations and the features of the Doppler
peaks in the
cosmic microwave background (see \cite{Lidseyetal} for a recent
review). 

There are many indications (summarized in \cite{PRD96}) 
that the inflaton couples explicitly to the Ricci curvature of
spacetime $R$, as described by the action
\be  
S= \int d^4 x \sqrt{-g} \left[  \frac{R}{2\kappa}-\frac{\xi
\phi^2 R}{2} - \frac{1}{2}  \nabla^{\mu} \phi
\nabla_{\mu} \phi  - V( \phi ) \right]  \; ,
\label{action}
\ee
where $\kappa \equiv 8\pi G $, $G$ is Newton's constant, $\phi $ the
inflaton field and $V( \phi )$ its potential.  Recently, much attention
has been paid to nonminimally coupled models of
inflation \cite{NMCinflation,FutamaseMaeda89,ALO}, dark matter
\cite{NMCdarkmatter}, and
quintessence \cite{NMCquintessence}. The slow-roll approximation that
allows to solve for the dynamics of inflation requires the
existence of 
inflationary attractors for the orbits of the solutions in phase space 
\cite{LiddleParsonsBarrow,SalopekBond}. While such
attractors are well known to exist for minimal coupling ($\xi=0$), their
presence is {\em assumed}, explicitly \cite{Kaiser} or tacitly, for
nonminimal coupling ($\xi \neq 0$).  Although slow-roll parameters have
been introduced, 
and  a Hubble slow-roll formalism developed, for $\xi \neq 0$
\cite{HwangCQG,HwangQ,Kaiser}, scenarios based on this
approximation would be empty theories if inflationary attractors in phase
space did not exist. In this Letter {\em general} potentials $V( \phi)$
and values of $\xi$ are considered and it is demonstrated that, subject to
the condition (\ref{21}), such attractors exist in the general 
nonminimally coupled case.

In a  spatially flat Friedmann-Robertson-Walker universe with line
element
\be   \label{metric}
ds^2=-dt^2 +a^2 (t) \left( dx^2 + dy^2 + dz^2 \right)  \; ,
\ee
the action (\ref{action}) yields the equations of motion
\be  \label{2}
6\left[ 1 -\xi \left( 1- 6\xi \right) \kappa \phi^2
\right] \left( \dot{H} +2H^2 \right) 
-\kappa \left( 6\xi -1 \right) \dot{\phi}^2   
- 4 \kappa V  + 6\kappa \xi \phi V' = 0 \; ,
\ee
\begin{equation}  \label{3}
\frac{\kappa}{2}\,\dot{\phi}^2 + 6\xi\kappa H\phi\dot{\phi}
- 3H^2 \left( 1-\kappa \xi \phi^2 \right) + \kappa  V =0 \, ,
\end{equation}
\be  \label{4}
\ddot{\phi}+3H\dot{\phi}+\xi R \phi +V' =0 \; ,
\ee
where an overdot and a prime denote, respectively, differentiation with
respect to the comoving time $t$ and to $\phi$. The phase space of the
system is two-dimensional \cite{2D,ALO} and $\left( H,
\phi \right)$ is a natural set of variables to describe the dynamics.
All the fixed points of the system (\ref{2})-(\ref{4}) are then
de Sitter solutions with constant scalar field,
\be  \label{dS}
\left( H, \phi \right) = \left( H_0, \phi_0 \right) \; ,
\ee
subject to the constraints
\be  \label{C1}
12 \xi H_0^2 \phi_0 +V_0' =0 \; , \;\;\;\;\;\;\;\;
H_0^2 \left( 1-\kappa \xi \phi_0^2 \right) = \frac{\kappa V_0}{3}  
\ee
(there are only two constraints since only two equations in the set
(\ref{2})-(\ref{4}) are independent).
For minimal coupling the well known solutions 
$\left( H, \phi \right) = \left( \pm \sqrt{\Lambda/3}, 0 \right) $ 
($\Lambda >0$) are recovered. In order to assess the stability of the
universes (\ref{dS}), one considers space- and time-dependent
perturbations,
\be 
H \left( t, \vec{x} \right) = H_0 + \delta H \left( t, \vec{x} \right) \;
, \;\;\;\;\;\;\;\;\;\;\; \phi \left( t, \vec{x} \right) =\phi_0 + \delta
\phi \left( t, \vec{x} \right) \; ;
\ee
unfortunately the perturbative analysis is plagued by the usual
gauge-dependence problems of general relativity. In a particular
gauge, one cannot be sure that the growing or decaying modes investigated  
are not pure gauge modes which can be removed by coordinate
transformations. We proceed by using the covariant and
gauge-invariant (GI) formalism of Bardeen \cite{Bardeen}, further  
developed in 
\cite{EllisBruni,HwangVishniac,MukhanovFeldmanBrandenberger};
a
recent version for generalized theories of gravity (including nonminimally
coupled scalar fields) was given in \cite{HwangCQG,Hwang}. We shall need
Bardeen's 
\cite{Bardeen} GI potentials $\Phi_H $ and $\Phi_A$ and the Ellis-Bruni
\cite{EllisBruni} variables
\be   \label{DeltaphiR}
\Delta \phi \left( t, \vec{x} \right) = \delta \phi + \frac{a}{k}\,
\dot{\phi} \left( B
- \frac{a}{k}
\, \dot{H}_T \right) \; , \;\;\;\;\;\;\;\;\;\;\;\;\;\;\;\;\;
\Delta R \left( t, \vec{x} \right) = \delta R + \frac{a}{k} \,\dot{R}
\left( B - \frac{a}{k} \, \dot{H}_T \right) \; ,
\ee
where $B$ and $H_T$ are metric perturbations and $k$ is  a
scalar harmonic eigenvalue \cite{Bardeen}. The evolution equations for the
variables $\Phi_{H,A}$ and $\Delta \phi$ were derived in \cite{HwangCQG}:
\be  \label{8}
\dot{\Phi}_H+\left( \frac{\xi \kappa \phi \dot{\phi}}{1-\kappa\xi\phi^2}
-H \right) \Phi_A 
-\frac{\kappa}{1-\kappa\xi\phi^2} \left \{ \xi \phi \Delta
\dot{\phi} +\left[ \xi \phi \left( \frac{ \dot{\phi}}{\phi} -H \right)
-\frac{\dot{\phi}}{2} \right] \Delta \phi \right\} =0 \; ,
\ee
\begin{eqnarray}  \label{9}
& & \left( \frac{k}{a} \right)^2 \Phi_H +\frac{1}{1-\kappa\xi\phi^2}
\left( \frac{3\xi^2\kappa\phi^2}{1-\kappa\xi\phi^2} + \frac{1}{2} \right)
\kappa \dot{\phi}^2 \, \Phi_A 
-\frac{1}{1-\kappa\xi \phi^2} 
\left\{ \left( \frac{3\xi^2\kappa \phi^2}{1-\kappa\xi\phi^2} + \frac{1}{2}
\right) \kappa \dot{\phi} \Delta \dot{\phi} \right. \nonumber \\
& & \left. +\left[ \left(  \frac{k}{a} \right)^2 \xi \phi 
-\ddot{\phi} 
\left( \frac{3\xi^2\kappa\phi^2}{1-\kappa\xi\phi^2} + \frac{1}{2} \right)
\right] \kappa \Delta \phi \right\} =0
\; ,
\end{eqnarray}
\be   \label{10}
\Phi_A + \Phi_H -\frac{2 \xi \kappa \phi \Delta \phi}{1-\kappa \xi \phi^2}
=0 \; ,
\ee
\begin{eqnarray}  \label{11}
& & \ddot{\Phi}_H +H \dot{\Phi}_H + \left( H- \frac{\xi \kappa \phi
\dot{\phi}}{1-\kappa\xi\phi^2} \right) \left( 2\dot{\Phi}_H -\dot{\Phi}_A
\right) -\frac{\kappa V }{1-\kappa \xi\phi^2} \, \Phi_A \nonumber \\ 
& & +\frac{\kappa}{1-\kappa\xi\phi^2} \left\{
- \xi\phi \Delta \ddot{\phi} +\left[ \frac{\dot{\phi}}{2} -2\xi \left(
\dot{\phi}+H\phi \right) \right] \Delta \dot{\phi} \right. \nonumber \\
& & \left. + \left[ \xi\phi \left( \kappa p-\frac{\ddot{\phi}}{\phi}
-\frac{2H\dot{\phi}}{\phi} \right) -\frac{V'}{2\kappa} \right]
\kappa \Delta \phi  \right\}=0 \; ,
\end{eqnarray}
\be  \label{12}
\Delta \ddot{\phi} + 3H \Delta \dot{\phi} + \left( \frac{k^2}{a^2}
 + \xi R +V'' \right) \Delta \phi + \dot{\phi} \left( 3
\dot{\Phi}_H -\dot{\Phi}_A \right) +2 \left( V' +\xi R \phi \right) \Phi_A
+ \xi \phi \Delta R =0 \; ,
\ee
where 
\be  \label{pressure}
p=\frac{1}{1-\kappa\xi\phi^2} \left[  \frac{\dot{\phi}^2}{2} -V -2\xi \phi
\left( \ddot{\phi}+3H\dot{\phi} + \frac{\dot{\phi}^2}{\phi} \right)
\right]
\; .
\ee
For the background universe (\ref{dS}) these yield, to first order in the
GI perturbations,
\be   \label{14}
\Phi_H=\Phi_A=\frac{\xi \kappa \phi_0}{1-\kappa\xi\phi_0^2} \, \Delta \phi
\; ,
\ee
\be  \label{15}
\Delta \ddot{\phi} + 3H_0 \Delta \dot{\phi} + \left[  
 \frac{k^2}{a^2}   +V_0''+
\frac{\xi R_0 \left( 1+\kappa\xi\phi_0^2 \right) +2V_0'
\kappa\xi\phi_0}{1-\kappa\xi\phi_0^2}  \right] \Delta \phi + 
\xi \phi_0 \Delta R =0 \; , 
\ee
plus a constraint equivalent to eqs.~(\ref{C1}); the
subscript zero denotes unperturbed
quantities. The GI variables $\Delta \phi$ and $\Delta R$ coincide
with the perturbations $\delta \phi $ and $\delta R$  to this order. The
expression
\be 
\Delta R =\delta R= \frac{-6\xi \kappa\phi_0 \left[ V_0''+4\left( 1+3\xi
\right) H_0^2 \right]}{1-\xi \left( 1-6\xi \right) \kappa\phi_0^2} \,
\Delta \phi \; , 
\ee
yields
\be  \label{17}
\Delta \ddot{\phi} + 3H_0 \Delta \dot{\phi} + 
\left( \frac{k^2}{a^2}   + \alpha \right) \Delta \phi  =0 \; , 
\ee
where 
\be  \label{alpha}
\alpha = \frac{V_0''\phi_0 \left( 1-\kappa\xi\phi_0^2 \right) -V_0' \left(
1-3\kappa \xi\phi_0^2 \right)}{\phi_0 \left[ 1-\xi \left( 1-6\xi
\right) \kappa \phi_0^2 \right]} \; .
\ee
Let us consider the expanding ($H_0 >0$) de Sitter solutions (\ref{dS}):
at late times $t \rightarrow + \infty$ one can neglect the 
$ \left( k/a \right)^2 \propto $e$^{-2H_0 t}$ term in eq.~(\ref{17}) and
look for
solutions of the form
\be    \label{19}
\Delta \phi \left( t, \vec{x} \right) = \frac{1}{(2\pi )^{3/2}} \int d^3
\vec{l} \,\, \Delta \phi_l (t) \, {\mbox e}^{i\, \vec{l} \cdot \vec{x} }
\;
,
\;\;\;\;\;\;\;   
\Delta \phi_l (t) =\epsilon_l \, \mbox{e}^{\beta_l \, t} 
\, ;
\ee
the constants $\beta_l$ must satisfy an algebraic equation with roots
\be
\beta_l^{( \pm )} =\frac{3H_0}{2} \left(  -1 \pm \sqrt{
1-\frac{4\alpha}{9H_0^2}} \, \right) \; .
\ee
Since {\em Re}$( \beta_l^{(-)} ) <0$ and the sign of 
{\em Re}$ ( \beta_l^{(+)} ) $ depends on $\alpha$, one
concludes
that the GI perturbations $\Delta \phi $ and $\Delta R \propto \Delta
\phi$ grow without bound unless
\be   \label{21}
V_0'' \geq f(x)\,  \frac{V_0'}{\phi_0}   \; ,
\ee
and that there is {\em instability}  otherwise; here
\be  \label{23}
x \equiv \kappa\xi\phi_0^2 \; , \;\;\;\;\;\;\;\;\;\;
f(x)=\frac{1-3x}{1-x} < 1 \; .
\ee
The stability condition (\ref{21}) is deduced by assuming that $0<x<1$ and
that
$\phi_0 \neq 0$ (the case $\phi_0=0$ is discussed later). If $x>1$ a
negative effective gravitational constant
$G_{eff} \equiv G \left( 1-\kappa\xi\phi_0^2 \right)^{-1} $ arises
\cite{FutamaseMaeda89}; moreover,  one of 
the slow-roll parameters (\ref{slowrollparameters}) diverges if the
solution $\phi (t)$ crosses the critical values 
$\pm \phi_1 \equiv \pm \left( \kappa\xi \right)^{-1/2} $ (for $\xi>0$) or 
$\pm \phi_2 \equiv \pm
\left[ \kappa\xi \left( 1-6\xi \right) \right]^{-1/2} $ (for $0< \xi <
1/6$). Under the  usual assumption that $V $ be non-negative, the
Hamiltonian
constraint
(\ref{3}) forces $| \phi|$ to be smaller than $\phi_2 $
\cite{FutamaseMaeda89,ALO}; we further assume that $| \phi | < \phi_1$ (if 
$ \left| \phi  \right| > \phi_1$ the direction of the inequality
(\ref{21}) is reversed).

In general, the stability of the fixed points (\ref{dS}) with 
$H_0>0$ depends on
the form of the potential $V( \phi ) $ and on the value of $\xi$.
However, the dependence from $\xi $ disappears and stability ensues
irrespective of the value of $\xi$ if\\
{\em i)} $V( \phi )$ has a minimum ($V'_0=0$ and $V_0''>0$) at
$\phi_0$ \\
{\em ii)} $V= m^2\phi^2/2$, or $\Lambda/\kappa + \lambda \phi^n$ 
($\Lambda, \lambda \geq 0$); the latter  is stable for $n \geq 1+f(x)$.

In the $\phi_0=0$ case not yet considered, eqs.~(\ref{8})-(\ref{12}) yield
\be   \label{24}
\Phi_H=\Phi_A=0 \; ,
\ee
\be  \label{25}
\Delta \ddot{\phi} +3H_0 \Delta \dot{\phi} + \left( \frac{k^2}{a^2}
 +\alpha_1 \right) \Delta \phi=0 \; ,
\ee
where $\Delta R =0$ and $\alpha_1= V_0''+4\xi \kappa V_0 $; hence there
is {\em stability} for $ V_0'' +4\xi\kappa V_0 \geq 0$ and instability
otherwise.

Finally, we consider the contracting ($H_0<0$) solutions (\ref{dS}); in
this case it is convenient to use conformal time $\eta$ (defined by $ dt
=ad\eta$) and the auxiliary variable $u \equiv a \Delta \phi$.
Eq.~(\ref{17}) becomes
\be   \label{29}
\frac{d^2u}{d\eta^2} + \left[ k^2 -U ( \eta ) \right] u=0
\; .
\ee
where
\be  \label{30}
U ( \eta ) = \left( 4-\frac{\alpha_1}{H_0^2} \right)
\frac{1}{\eta^2}+\frac{2}{H_0 \eta^3} \; ,
\ee
and we used the relation $\eta =- ( aH_0 )^{-1} $ valid in the background
(\ref{4}). Formally, eq.~(\ref{29}) is a one-dimensional Schr\"{o}dinger
equation for a quantum particle of unit mass in the potential $U( \eta )$;
its
asymptotic solutions  at large $\eta $ (i.e. $t\rightarrow + \infty$) are 
free waves $u \simeq {\mbox e}^{\pm i \, k \eta}$, and $\Delta \phi 
\propto H_0 \eta  $ diverges. The solutions (\ref{dS}) with $H_0<0$ are
{\em unstable}, as in the $\xi=0 $ case.

The case $\phi = \pm \phi_1 $ not considered above corresponds to a class
of solutions with constant Ricci curvature containing a de Sitter
representative (\ref{dS}); however the latter is clearly fine-tuned and
unstable with respect to perturbations $\Delta \phi$.

As a conclusion, subject to the condition (\ref{21}),  
inflationary attractors exist in nonminimally, as well as in 
minimally, coupled inflation and the slow-roll approximation for $\xi \neq
0$ is meaningful. The Hubble slow-roll
parameters (\cite{HwangCQG}-see also \cite{Kaiser}) 
\be  \label{slowrollparameters}
\epsilon_1 = \frac{\dot{H}}{H^2} \; , \;\;\;\;
\epsilon_2 =\frac{\ddot{\phi}}{H \dot{\phi}} \; , \;\;\;\;
\epsilon_3 =-\, \frac{\xi \kappa \phi \dot{\phi}}{H \left[ 1- \left(
\frac{\phi}{\phi_1} \right)^2 \right]} \; , \;\;\;\;
\epsilon_4 =-\, \frac{\xi \left( 1-6\xi \right) \kappa \phi \dot{\phi}}
{H \left[ 1-\left( \frac{\phi}{\phi_2} \right)^2 \right] } \; ,
\ee
vanish {\em exactly} for the solutions (\ref{dS}) ($\epsilon_4$ also 
vanishes for conformal coupling $\xi=1/6$), and $\left| \epsilon_i \right|
<< 1$ for the solutions attracted by (\ref{dS}) in the phase 
space\footnote{The importance of the inflationary attractors is made
clear by the fact that, for the {\em contracting} solutions (\ref{dS}),
the slow-roll approximation is exact (in the sense that the slow-roll
parameters (\ref{slowrollparameters}) vanish); however, this bears no
relationship with the actual inflationary solutions because the
contracting spaces (\ref{dS}) are not attractors.}.

The spectra of scalar and tensor perturbations in nonminimally
coupled inflation were calculated in \cite{HwangQ}
as special cases of more general gravity theories, and the respective
spectral indices are
\be
n_S=1 +2 \left( 2 \epsilon_1 - \epsilon_2 + \epsilon_3 -\epsilon_4
\right)  \; , \;\;\;\;\;\;\;\;\;\;\;\;\;
n_T=2 \left( \epsilon_1 - \epsilon_3 \right) 
\ee
computed at the time when the perturbations cross outside the horizon 
(these formulas can be found in \cite{Kaiser}); applications to specific
inflationary scenarios with nonminimal coupling will be given elsewhere.

\section*{Acknowledgments}

This work was supported by the EEC grant PSS*~0992 and by OLAM, Fondation
pour la Recherche Fondamentale, Brussels.

\end{document}